\documentclass[journal]{IEEEtran}
\usepackage{cite}
\ifCLASSINFOpdf
\else
\fi
\usepackage{hyperref}
\usepackage{array} 
\usepackage{mdwmath}
\usepackage{graphicx}
\usepackage{amsfonts}
\usepackage{tabularx}
\usepackage{cite}
\usepackage{amsmath}
\makeatletter
\def\BState{\State\hskip-\ALG@thistlm}
\makeatother
\usepackage{xcolor}
\usepackage{mathtools}

\usepackage{hyperref}
\usepackage{epstopdf} % When using eps figures

\usepackage{fancyhdr}
\fancypagestyle{firstpage}
{
    \fancyhead[L]{}    
    \fancyhead[R]{}
}
\begin{document}

\title{IoT Coverage Enhancement using Repetition in Energy Constrained Devices: an Analytic Approach}

\author{\IEEEauthorblockN {Bisma Manzoor,~\IEEEmembership{Student Member,~IEEE}, Bassel Al Homssi,~\IEEEmembership{Member,~IEEE},\\ and Akram Al-Hourani~\IEEEmembership{Senior Member,~IEEE}}
\thanks{The authors are with the School of Engineering, RMIT University, Melbourne, VIC 3000, Australia. This research is supported by the Australian Government Research Training Program (RTP) Scholarship.}
}
\maketitle
%Hence, novel access technologies have emerged that consider the specific requirements for IoT communications.
% These transmissions are combined at the receiver

%Applications relying on wireless Internet of Things (IoT) have significantly increased over the past decade. Typical broadband cellular technologies are not intended for the massive machine-type IoT communications that rely on energy-limited devices. 
%uch as NB-IoT,

\begin{abstract}
\thispagestyle{firstpage}
\lhead{\small{ }}
Novel Internet-of-Things (IoT) access technologies are emerging as part of the next generation cellular networks. These technologies are specifically oriented towards energy-limited IoT devices that are scattered far away from their serving base station. One of the key methods of achieving deep coverage is via repeated transmissions of data. However, repetition leads to higher energy consumption and reduces the device battery-lifetime. Thus, a trade-off between coverage and energy consumption exists and requires careful investigation. This paper evaluates the effects of transmission repetition on enhancing the probability of coverage and the cost of the incurred energy. Using an empirical repetition
profile based on traffic load and device distance from the base station, we derive the coverage probability and energy profile models for IoT links that utilize two different diversity combining techniques. In particular, we focus on two common diversity combining that are; (i) Selection Combining (SC), and (ii) Maximal Ratio Combining (MRC). We utilize tools from stochastic geometry to formulate an analytic framework that compares these two combining methods. This framework can aid network designers in jointly maximizing the network coverage while minimizing the energy expenditure of devices.
\end{abstract}


\section{Introduction}
Cellular Internet of Things (CIoT) connections are expected to grow to 5 billion by 2025 \cite{RepExerricson} due to the increasing number of use-cases from various sectors such as agriculture, healthcare, industry. The 5G New Radio (NR) is provisioned to support the growing traffic in the International Mobile Telecommunications (IMT) network, that also includes CIoT. The use cases of 5G are broadly classified into three categories (i) \textit{massive Machine-Type Communication (mMTC)}, (ii) \textit{ultra-Reliable Low Latency Communication (uRLLC)}, and (iii) \textit{enhanced Mobile Broadband (eMBB)}\cite{RepExitu}. IoT applications are envisioned to fall under mMTC and uRLLC. mMTC supports discontinuous and delay-tolerant applications, whereas uRLLC is meant for critical and high throughput applications\cite{RepExintro1}. 

The massive number of connections in mMTC is due to the wide range of applications across smart cities, smart agriculture, and others\cite{RepExsmartcity}. mMTC IoT devices are expected to have a longer battery life compared to the traditional cellular devices, and operate at lower power and long ranges.
A class of network technologies designed to support low-power IoT devices is termed as Low-Power-Wide-Area-Networks (LPWAN) which provides wireless communication services while meeting the service requirements of wide area coverage, and low
battery consumption for mMTC devices at the cost of lower bit rate than typical cellular services.

Narrow band Internet of Things (NB-IoT) is one of the LPWAN technologies recently developed by 3GPP to operate in the licensed frequency spectrum. Unlike mobile broadband, NB-IoT supports low-power and high coverage by maintaining low bit-rate. There is a multitude of methods available to enhance the coverage, such as utilizing narrow-bandwidth, using spread spectrum techniques, and employing frame repetitions~\cite{RepExbook2}. Frame repetition is gaining traction in new IoT access technologies such as NB-IoT. It is also utilized in uRLLC for grant-free access mechanism~\cite{RepExkrep} that allows a device to access time-frequency resources without going through the traditional handshake process.
Repetitions allow the data to be transmitted multiple times, increasing the detection-ability by enhancing the signal power at the receiver. However, it comes at the cost of increased number of transmissions resulting in an overall increase in network traffic and interference. In addition, even though repetitions improve the coverage, they cause an extensive battery consumption of the transmitting device. This is due to the transmitter staying active for a prolonged duration while transmitting the repeated data. Furthermore, each unsuccessful transmission leads to energy wastage in the transmitting device. Thus, it is imperative to jointly analyze the coverage and energy wastage of the network. Besides, various combining techniques exist to process the repeated transmissions at the receiver. The diversity combiners operate at different stages of the receiver. The repetitions are either combined before signal processing which leads to an increase in total signal energy, or the combination takes place after signal processing, resulting in an overall improvement in the system performance\cite{RepExdiversity5}. 

A plethora of good literature is available on the energy consumption of IoT devices~\cite{RepExenergy3,RepExenergy4,RepExenergy1,RepExenergy2}. However, the energy wasted due to the failure of repetitions does not seem to be analyzed previously. Besides, authors in \cite{RepExMRC2,RepExMRC3} present the analysis on combining techniques, and the work in \cite{RepExpap1} sheds light on the combination scheme for IoT; however, the impact of interference has not been taken into account. This work focuses on analyzing the repetition scheme for the combining techniques that operate before signal processing which are Selection Combining and Maximal Ratio Combining. The coverage probability and energy dissipation analysis are carried out for both diversity combining techniques. The performance is examined under the influence of spatially co-related interference. The contributions of this paper are summarized as follows,
\begin{itemize}
\item It provides analytical models for the coverage probability with repetitions using \textit{SC} and \textit{MRC}.
\item A mathematical model has been put forth to capture the energy expenditure of a device during the process of implementing repetition.
\item Furthermore, the model aids in obtaining the ideal repetition profiles that facilitate the maximum service availability and minimum energy dissipation. 
\end{itemize}
\section{Background and Related Work}
 The concept of repetition is introduced by 3GPP~\cite{RepEx3gpp} in release 13, for cellular IoT access technologies to extend the coverage in mMTC networks comprising of vast number of devices~\cite{RepEx2D}. The devices using the repetition scheme repeat the frames multiple times, which are accumulated at the receiver to enhance the signal detection probability. The signal is then processed for synchronization and channel estimation~\cite{RepExbook2}.
 Ideally, the processing gain at the receiver should be proportional to the number of repetitions used. Nevertheless, practical signal combination is imperfect due to channel impairments resulting in a lower gain than that of an ideal gain.

Signal propagation in a channel is subject to power losses due to multiple factors including the losses due to natural wave expansion and the losses due to multi-path fading. The average attenuation of the signal power as it propagates for a certain distance is termed as the \textit{path-loss} while the fluctuations in the signal power due to the interactions with the environment are caused by the multipath signal components, this is referred to as \textit{fading}~\cite{RepExitureport}. In a non-stationary channel, the different repetitions encounter different levels of fading~\cite{RepExNbIoT}, which is mitigated by combining the repetitions at the receiver. As different combination techniques exist~\cite{RepExDC}, their performance analysis under variable signal impairments is of interest. The most commonly used diversity combining methods are (i) \textit{SC} (ii) \textit{MRC}. In SC technique, the receiver opts for the signal with the highest SNR. MRC is a linear combining technique, where, all the signals are weighted in terms of amplitude and phase, followed by the combination at the receiver, leading to the gain in the signal power~\cite{RepExdiversity4}. 
 Although the performance is improved with the aid of combining techniques, it is vital to investigate the impact of interference on the received signals.

One of the main challenges to accurately examine network performance is to incorporate the channel losses and the interference from the other active devices located randomly in the network. Stochastic geometry captures the spatial randomness of IoT devices and thus facilitates the analysis of key performance indicators in a wireless network~\cite{RepExASInterference}. Using the point process to define the location of the devices usually makes the analytic approach more tractable~\cite{RepExMartin1}. Literature overview on combining techniques using stochastic geometry is presented in~\cite{RepExpap2,RepExDP2}, where in the work in~\cite{RepExpap2} provides the performance analysis of the SC and MRC techniques using the Nakagami-Fading Channels, while in~\cite{ RepExDP2}  the authors provide the quantitative measures of the performance of the linear combining techniques including SC and MRC, along with the analytical expressions for SNR associated with the given combining techniques are put forth. Nevertheless, the impact of unwanted power arising from the spatially distributed nodes is not accounted for. Moreover, the work in~\cite{RepExpap1} provides a performance analysis of the combining strategies implemented in cellular IoT technologies that are SC and MRC; however, it does not account for the interference in the network. In contrast, although the authors in~\cite{RepExMartin,RepExMRC}, employ stochastic geometry analysis with interference correlation, which aids in profoundly characterizing the network behavior for various combination techniques, there is a lack of investigation for the IoT repetition scheme under such scenario.

Additionally, the repetition scheme impacts the active period of the device leading to more energy consumption. An IoT device is said to be in an active state when it is communicating with the network, else it is inactive in order to preserve the battery~\cite{RepExenergy1}. To further enhance the battery life in IoT devices, operational modes, i.e., Power Saving Mode (PSM) and extended Discontinuous Reception (eDRX), are employed in cellular IoT technologies. However, since the repetition scheme is employed in the active state, the impact of repetitions on energy drainage during this state requires investigation, especially if the repetitions fail to get delivered. Various innovative mechanisms have been adopted to enhance the battery efficiency in IoT devices. One such technique is the energy harvesting~\cite{EH1}, where the IoT devices harvest their energy from the external sources. The work in~\cite{EH1}, have put forth a  two stage algorithm  to maximize the energy efficiency in IoT devices by harvesting energy from the radio-frequency signals. Authors in~\cite{EH2} use the energy harvesting mechanism for  an access scheme called irregular repetition ALOHA (IRA) that utilizes varying repetitions. They put forth the optimal transmission policy to maximize the system throughput. In another work~\cite{EH3} authors have proposed a Irregular Repetition Slotted ALOHA, where the nodes replicate the packets on a frame while utilizing the energy harvesting. The authors obtain the optimized probability distributions for the number of packet replicas. However, as aim of repetitions employed in the cellular IoT is to achieve a high SNR and the detection probability at the receiver~\cite{ruki}, the above literature lacks the analysis of the energy efficiency from such perspective. Furthermore, the literature on energy-efficient techniques involving the repetitions in IoT devices is analyzed in~\cite{RepExenergy3, RepExenergy4,RepExenergy1}, wherein the predictive algorithm proposed in~\cite{RepExenergy4} aims at lowering energy expenditure due to repetitions in an NB-IoT network by reducing the scheduling requests. 
The work in~\cite{RepExenergy2} has highlighted the impact of repetitions on energy consumption and hence, focused on the significance of implementing appropriate network parameters, such as repetition rate. Nevertheless, the above works do not investigate the energy wastage due to unsuccessful repetitions. \\
 
\section{System Model}\label{sysmodel}
%In this section, the system model classified into multiple subsections is presented. It comprises of geometric distribution of IoT devices, the repetition model and the channel model between the IoT devices and the gNB. 

\subsection{Geometric Model}\label{geomodel}
\begin{figure}
    \includegraphics[width=\linewidth]{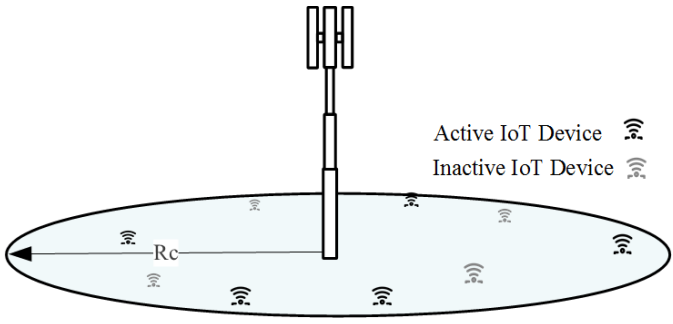}
        \caption{Illustration of a typical IoT cell.}
    \label{fig:fig 1re}
\end{figure}
We consider a cell with a radius $R_\text{c}$ and a gNB located at its center as illustrated in Fig.\ref{fig:fig 1re}. The IoT devices are assumed to be distributed by an isotropic and homogeneous Poisson Point Process (PPP), denoted by $\Phi \in \mathbb{R}^2$. The active devices have an initial \textit{base density} of $\lambda_\text{o}$ and are assumed to transmit messages of equal length at a constant power $P_\text{t}$.
\begin{table}[t]\label{tab1}
\centering
\caption{Notations and Symbols }
\begin{tabular}{p{1cm}p{5cm}p{1cm}}
\hline
\hline
\bf Symbol & \bf Definition & \bf Value\\
\hline
$\Phi$ & IoT devices point process&-\\
$\lambda_\text{o}$& Initial base density & 2$/100^{2}$ \\
$D_\text{o}$& Initial base duty cycle& 0.01\\
$r_\text{o}$& Distance of device from a gNB&-\\
$\lambda(r)$& Effective density&-\\
$D(r)$& Effective duty cycle&-\\
$N$& Repetitions&-\\
$\psi(r)$& Repetition profile function&-\\
$a$& Steepness parameter of repetition profile&-\\
$b$& Rate parameter of repetition profile&-\\
$F_I$& CDF of the Interference&-\\
$R_\text{c}$& Radius of the cell&1000~m\\
$p|_{r{_\mathrm{o}}}(1)$& Probability of success for a single transmission given the device is at distance $r_\mathrm{o}$&-\\
$p|_{r{_\mathrm{o}}}(N)$& Probability of success for $N$ repetitions given the device is at distance $r_\mathrm{o}$&-\\
$\epsilon(1)$& Energy consumed for a single transmission&-\\
$\epsilon({N})$& Energy consumed for $N$ transmissions&-\\
$\epsilon_\mathrm{w}(N)$ & Energy wasted for $N$ transmissions&-\\
$\Bar{\epsilon}_\text{c}$& Average energy wasted in a cell&-\\
$f_r(r_\mathrm{o})$ & Radial distance pdf&-\\
B & Bandwidth & 180~kHz\\
F & Receiver noise figure & 3~dB\\ 
$\eta_\epsilon$ & Power conversion factor&4\\
$P_\text{O}$ & Power overhead &210~mW\\
$\alpha$ & Path-loss exponent & 3.5$^{\dagger1}$\\
$\sigma^2$ & Average noise power & $-$118~dBm$^{\dagger2}$\\
\hline
\multicolumn{3}{l}{$\dagger1$~\footnotesize{WINNER II channel model measurements~\cite{RepExwinner}}}\\
\multicolumn{3}{l}{$\dagger2$~\footnotesize{Calculated from noise figure of 3 dB and bandwidth 180~kHz}}
\end{tabular}
\end{table}

To enhance the signal coverage in this IoT network, we consider the repetition mechanism, according to which the entire message is repeated several times\cite{RepEx3gpp}. 
We utilize the mathematical framework
developed in \cite{RepExRepPaper} in order to implement the repetition profile that is based on the duty cycle and the device's distance from gNB. We assume $D_\text{o}$ is the base duty cycle in a cell, given as $D_\text{o}=T_\text{o}/T_\text{max}$, where $T_o$ is the duration of a single radio frame that is sent after every $T_\text{max}$ time. Since the devices are configured to repeat $N$ times, the effective duty cycle becomes $D = NT_\mathrm{o}/T_\mathrm{max} = N D_{\mathrm{o}}$.

\begin{figure}[t]
    \includegraphics[width=\linewidth]{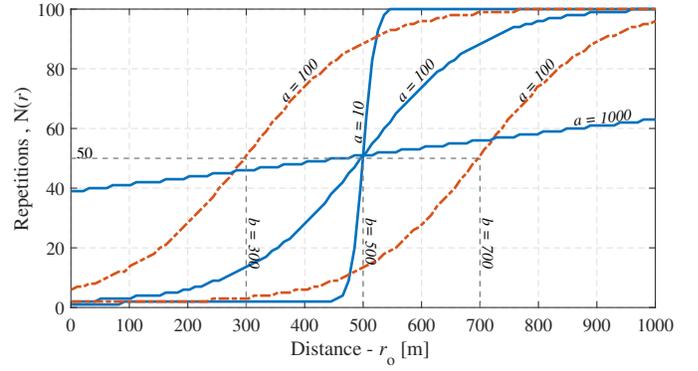}
        \caption{Repetitions $N(r)$ versus the radial distance at various repetition profile parameters.}
    \label{fig:rep1}
\end{figure}

Moreover, the signal strength weakens for the devices located father from gNB, and thus they tend to repeat more. This intuitively increases the duty cycle of the devices. To incorporate this impact of distance, the duty cycle as a function of distance is put forth as,
\begin{equation} \label{eq1}
D(r) = D_\mathrm{o} + (1-D_\mathrm{o})\psi(r)
\end{equation}
Hence, using the relation between the duty cycle and $N$ as previously put forth, the repetition profile in terms of distance is given as follows,
\begin{equation}\label{eq2re}
N(r)= \left\lceil{1 + \frac{1-D_{\mathrm{o}}}{D_{\mathrm{o}}} \psi(r)}\right\rceil,
\end{equation}
where $\lceil.\rceil$ is the ceiling function, $D_\text{o}\leq D(r)\leq1$, and $\psi(r)$ is called the repetition profile function. This function is selected such that $\psi(r)\in\left[0,1\right]$, and the duty cycle is bounded between 0 and 1, i.e., $D(r)\in\left[0,1\right]$, since duty cycle cannot exceed unity. Also $\psi(r)$ should be monotonically increasing, such that $N(r)$ increases with the distance. A suitable function that satisfies the above requirements is the logistic function given as,
\begin{equation}
    \psi(r)=\frac{1}{1+\exp\left(-\frac{r-b}{a}\right)},
\end{equation}
where $a$ and $b$ are the repetition profile function parameters of the repetition profile. $a$ controls the steepness of repetitions, while, $b$ determines the mid-point of the logistic function. Repetition profiles with a varying $N$ obtained at various repetition parameters is illustrated in Fig.~\ref{fig:rep1}. It can be seen that the number of repetitions used by the device increases with respect to distance from gNB, and every repetition profile corresponds to different number of repetitions. The model was also empirically observed in NB-IoT technology~\cite{RepExNbIoT}. Moreover, by utilizing the above given model, the density at a given snapshot in a cell using a certain repetition profile is defined as effective density, denoted as $\lambda(r)$, and is given as $\lambda(r) = \lambda_\mathrm{o}D(r)$.

\subsection{Channel Model}
For a device located at $x$, the received power at the gNB located at the origin is $P_r= P_{\mathrm{t}} h~ l(x)$, where $h$ is the channel fading random variable and $l(x)$ is the path-gain\cite{OPTZpathloss}. Using the log-distance path-loss model, $l(x)=||x||^{-\alpha}$~\cite{RepExlogdistance}. 
The strength of the signal at the receiver is also subject to additional channel losses such as shadowing; however, this work focuses on evaluating the performance under fading and path-loss only. Furthermore, for various combining techniques, the success probability at the receiver, which is dependent on the received signal power and the interference from other devices, is analyzed for two-channel conditions (i) \textit{Path-loss only} where $h=1$ and the loss in the signal is only due to the distance of the device from the gNB. (ii) \textit{Path-loss with Rayleigh fading}, which includes the path-loss along with the Rayleigh fading, where $h$ is assumed to be an exponentially distributed random variable with unity mean. We first evaluate the interference characterization for both channel conditions, followed by performance evaluation.  

\section{Interference Characterization}\label{IC}
To preserve the simplicity of the IoT access technology, we assume a simple ALOHA-like system where radio frames collide due to overlapping transmissions. For a device being served at location $||x_\text{o}||$, all the other active users are considered as a source of interference. In order to evaluate the success probability, it is essential to characterize the aggregate interference power from the active users as follows, 
\begin{align}\label{agI}
I = \sum_{x_\text{i}\in\Phi \text{\textbackslash} x_{\mathrm{o}}} P_{\mathrm{t}} h_\mathrm{i}||x_\mathrm{i}||^{-\alpha}=\sum_{x_\mathrm{i}\in\Phi \text{\textbackslash} x_{\mathrm{o}}} P_{\mathrm{t}} h_\mathrm{i}r_\mathrm{i}^{-\alpha}, 
\end{align}
where $||x_i||$ is the location of the interferers located at a distance $r_i$ from the gNB. 
The stochastic behavior of the interference is typically characterized using its cumulative distribution function (CDF), however, such function cannot be obtained in a closed-form for the proposed scenario. Alternatively, we describe the stochastic behavior of the interference in terms of the Laplace Transform (LT) of the CDF. For a known LT of the interference, the CDF is calculated using numerical inversion of Laplace transform~\cite{OPTZ8770117},
\begin{align}\label{eq6}
    F_I(x)=\mathcal{L}^{-1}\left[\frac{1}{s}\mathcal{L}_I(s)\right],
\end{align} 
where $\mathcal{L}_I(s)$ is the LT of the interference random variable $I$ evaluated at a point $s$ which is defined as ${\mathcal{L}_X(s)= \mathbb{E}[{\mathrm{exp}(-sX)}]}$, for a given variable $X$.
Using the above definition of Laplace transform, $\mathcal{L}_I(s)$ is evaluated for $I$ from \eqref{agI}, as follows,
\begin{align}\label{eq7re}
 \mathcal{L}_I(s) 
    &= \mathbb{E}_{\Phi}\mathbb{E}_{h}\left[\mathrm{exp}\left(-s\sum_{x_i\in \Phi\setminus x_\mathrm{o}}h_i r_i^{-\alpha}\right)\right]\nonumber\\
    &\overset{(\mathrm{a})}= \mathbb{E}_{\Phi}\left[\prod_{x_i\in \Phi\setminus x_\mathrm{o}}\mathbb{E}_{h}\left[\exp\left(-s h_i r_i^{-\alpha}\right)\right]\right],
\end{align}
where (a) stems from the fact that $h_i$ is independent of the point process and by converting the sum of exponents to product. Using the definition of the probability generating functional on $\mathbb{R}^2$, for a function $f : \mathbb{R}^2 \rightarrow [0, 1]$ as in~\cite{OPTZ8770117}, 
\begin{align} \label{pgfl}
 \mathbb{E}\left[\prod_{x\in \Phi}f(x)\right]= \mathrm{exp}\left(-\int_{\mathbb{R}^2}\left[1-f(x)\right]\lambda(\mathrm{d}x)\right)  
\end{align}
for the PPP, $\mathcal{L}_I(s)$ is given as, 
\begin{align}\label{lapgeneral}
   \mathcal{L}_I(s) =\mathrm{exp}\left(-\int_{\mathbb{R}^2}
   \left(1-\mathbb{E}_{h}\left[\exp\left(-s h x^{-\alpha}\right)\right]\right)\lambda(\mathrm{d}x)\right)~.
\end{align}
\subsection{Interference Characterization with path-loss only}
Considering the scenario for the channel when it is impacted only by path-loss. Thus, ignoring the effect of fading and substituting the fading component $h=1$, we re-write~\eqref{lapgeneral} in polar coordinates as follows, 
\begin{align}\label{eq8re}
   \mathcal{L}_I(s)=
     \exp\left(-2\pi\lambda_\mathrm{o}\int_{o}^{R_\text{c}}\left[1-\exp\left(-sr^{-\alpha}\right)\right] D(r) r\mathrm{d}r\right).
\end{align}
Thus, the inverse Laplace Transform of \eqref{eq8re} provides the CDF of the aggregate interference power and it can be numerically obtained using the Euler algorithms for numerical approximations of the inverse Laplace Transform~\cite{RepExInverse}.

\subsection{Interference Characterization with Rayleigh Fading}
In this scenario we assume that both the serving and the interfering signals are impaired with path-loss and Rayleigh fading, which is denoted by an exponential distribution random variable with a unity mean. The definition of the moment generating function (MGF) of such an exponential variable is ${E}\left[\exp(-t)\right]=1/(1-t)$. Thus MGF for the fading variable $h$ at $-sr_\text{i}^{-\alpha}$ is given as, $\mathbb{E}_{h}\left[\exp(-shr_{i}^{-\alpha})\right]=1/(1+sr_i^{-\alpha})$
Substituting in \eqref{lapgeneral}, the Laplace transform of interference for Rayleigh fading is given as in~ \cite{RepExRepPaper},
\begin{align}\label{eq9are}
     \mathcal{L}_{I}(s)= \exp\left(-2\pi\lambda_\mathrm{o}\int_{0}^{R_\mathrm{c}}\frac{s}{s+r^{\alpha}}D(r)r\mathrm{d}r\right).
\end{align}
Consequently,~\eqref{eq8re} and~\eqref{eq9are} are employed to calculate the success probability in terms of coverage for the given combining techniques utilizing repetition model. In case of the scenario when no repetitions are used in the cell, $\psi(r)=0$, and $\lambda(r)$ reduces to $\lambda_\mathrm{o}$ in \eqref{eq8re} and \eqref{eq9are}. 

\section{Performance Evaluation}
This section introduces the performance evaluation of the two diversity combining techniques (i) \textit{SC} and (ii) \textit{MRC} at the given repetition profiles compared to the no repetition scenario. The values of the repetition profiles parameters $a$ and $b$ can be optimized based on the evaluation of repetition performance zones as defined in~\cite{RepExRepPaper}. The combining techniques are investigated under the spatially-correlated interference and channel impairments that include path-loss and Rayleigh fading. 
The probability of success at the receiver is assessed in terms of the Signal-to-Interference-plus-Noise-Ratio (SINR), denoted by $\gamma$. It is the ratio of the received signal power of the target device to the sum of total interference power from interferes and noise power, given as $\gamma=h_\mathrm{o} r_\mathrm{o}^{-\alpha} /(I+\sigma^2)$, where $r_\mathrm{o}=||x_\mathrm{o}||$ is the distance of the target device from the gNB, $\sigma^2$ is the average noise power, $I$ is the aggregated interference power from the active interferes. A transmission is considered to be successful when $\gamma$ is above a given design threshold $\theta$ at the receiver. 
\subsection{No Repetition}\label{NoRep}
This is a typical scenario where the devices are configured to transmit the message only once. We abbreviate it as \textit{No Rep} and use the same notation for comparison. The probability of having the SINR of this single transmission above a required receiver threshold is formulated as,
  \begin{align}\label{norep}
      p_{\mathrm{\tiny{No Rep}}}= \mathbb{P}\left(\frac{h_\mathrm{o} r_\mathrm{o}^{-\alpha} }{I+\sigma^2}>\theta\right)=F_I\left(\frac{h_\mathrm{o}r_\mathrm{o}^{-\alpha}}{\theta}-\sigma^2\right),
  \end{align}
The above is considered as the probability of success and is later compared with the combining techniques.
For path-loss only, the probability of success is obtained by evaluating $F_I$ using the inversion of Laplace transform from \eqref{eq8re} and by substituting ${D(r)=D_o}$. In case of the channel impaired by fading, we assume both Rayleigh fading and the path-loss. Thus, the probability of success is calculated with the Rayleigh fading component $h_\mathrm{o}$ as an exponential distribution random variable with unity mean. Further, $F_I$ in this case is evaluated using the inversion of Laplace transform from \eqref{eq9are}, with $D(r)=D_o$, and is deconditioned over the fading by utilizing the pdf of exponential distribution random variable given by ${f_{h_\mathrm{o}}(x)=\exp(-x)}$. The concept is well defined in literature as given in \cite{RepExASInterference}. 
 
\subsection{Selection Combining}\label{sc}
In this combination technique, at a given time, the receiver selects the repetition with the highest SINR \cite{RepExDP2}. This is expressed as,
\begin{align*}
    \gamma_\text{sc}=\max(\gamma_\text{1},\gamma_\text{2}...\gamma_\text{N}),
\end{align*}
where, $\gamma_\text{1},\gamma_\text{2},..\gamma_\text{N}$ are the individual SINRs of the repetitions 1 to N. Accordingly, in order to have a successful repetition, there should be at least one signal with SINR above the required threshold. Thus, the probability of success conditioned at a given $r_\mathrm{o}$ is formulated as, 
\begin{align}\label{sc1}
    p|_{r{_\text{o}}}(N)=\mathbb{P}\left(\exists~n \in [1,N]: \gamma_\text{n} > \theta \right)
\end{align}
 Since the receiver selects the signal with the highest SINR, the probability of the chosen signal to be below a given threshold is the product of probabilities of SINRs of all signals to be below the threshold\cite{RepExbook3,RepExmolish} accordingly the, probability of error is,
\begin{align}\label{pe}
   p_e{_{|r_\mathrm{o}}}(N)=\mathbb{P}\left[ \gamma_\text{n}<\theta\right]^N=\left[1-p|_{r{_\text{o}}}(1)\right]^N,
\end{align}

where $p|_{r{_\text{o}}}(1)$ is the probability of success of a single signal and $n\in[1,N]$. Therefore, the probability of success for SC is given by 
\begin{align}\label{SCnew}
   {p_{|r_\mathrm{o}}(N)}&=1-p_e{_{|r_\mathrm{o}}}(N) 
\end{align}
which is evaluated for two channel scenarios as described next.

\subsubsection{SC with path-loss only}\label{scPL}
The analysis in this case presents the upper 
bound of the channel performance since both the serving and interfering signals are impacted by path-loss only channel. Thus, the probability of success for a signal at $r_o$ is, 
\begin{align}\label{eq5re}
    p_{|r_\mathrm{o}}(1)=\mathbb{P}({\gamma >  \theta}|r_\mathrm{o}) 
    &=F_I\left(\frac{r_\mathrm{o}^{-\alpha}}{\theta}-\sigma^2\right),
\end{align}
where $F_I$ is the (CDF) of the interference $I$ and is obtained using~\eqref{eq6} and~\eqref{eq8re}. Hence, given $N$ repetitions the success probability is evaluated using~\eqref{SCnew} as follows,
\begin{align}\label{eqSCpathloss}
       {p_{|r_\mathrm{o}}(N)}&=1-p_e{_{|r_\mathrm{o}}}(N)\nonumber\\&=1-\left[1 - F_I\left(\frac{r_\mathrm{o}^{-\alpha}}{\theta}-\sigma^2\right)\right]^N,
\end{align} 
\subsubsection{SC with Path-loss and Fading}\label{sc2}
We take the Rayleigh fading model which represents the worst-case scenario with no dominant path between the device and the gNB. Thus, the probability of success at the receiver conditioned at a given $r_o$ and $h_o$ using~\eqref{pe} is, \begin{align}\label{scfading}
      p_{|r_\mathrm{o},h_\text{o}}(1)&=\mathbb{P}(\gamma >  \theta|r_\mathrm{o},h_\text{o}) \nonumber\\&=   \mathbb{P}\left(\frac{h_o r_\mathrm{o}^{-\alpha} }{I+\sigma^2}>\theta\right)\nonumber\\
       &=F_{I}\left(\frac{h_\mathrm{o}r_\mathrm{o}^{-\alpha}}{\theta}-\sigma^2\right),
\end{align}
where $h_o$ is independent of $r_\mathrm{o}$. Deconditioning at $h_\mathrm{o}$ , 
\begin{align}\label{}
      p_{|r_\mathrm{o}}(1)=\int_0^\infty F_{I}\left(\frac{u r_\mathrm{o}^{-\alpha}}{\theta}-\sigma^2\right) f_{h_{\mathrm{o}}}(u)\mathrm{d}u,
\end{align}
Thus, the probability of success for N repetitions, at $r_\mathrm{o}$ is obtained using~\eqref{SCnew} as follows, 
\begin{equation}\label{eq4re}
\begin{split}
       {p_{|r_\mathrm{o}}(N)}&=1-p_e{_{|r_\mathrm{o}}}(N)\\
       &=1-\left[1 - \int_0^\infty F_{I}\left(\frac{u r_\mathrm{o}^{-\alpha}}{\theta}-\sigma^2\right) f_{h_{\mathrm{o}}}(u)\mathrm{d}u\right]^N,
\end{split}
\end{equation} 
where the CDF of the interference in this case is obtained by using~\eqref{eq6} and~\eqref{eq9are}. $f_{h_\text{o}}(u)=\exp(-u)
$ is the PDF of the exponential distribution with unity mean.

\begin{figure}[t]
    \includegraphics[width=\linewidth]{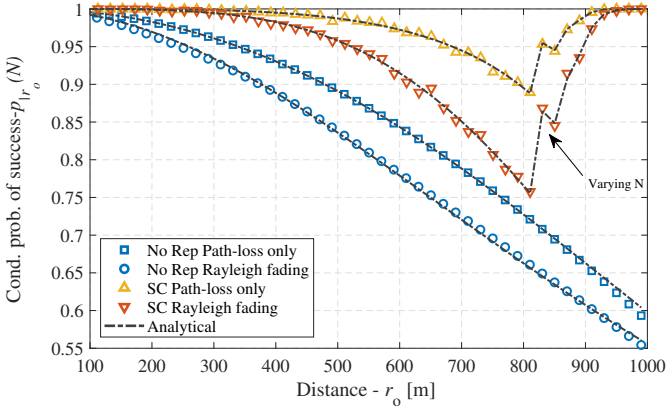}
        \caption{Conditional success probability vs $r_\mathrm{o}$ obtained for no repetition and the two cases of Selection Combining from~\eqref{eqSCpathloss} and \eqref{eq4re}, respectively. The utilized repetition profile parameters are $a = 50$, $b=1050$.}
    \label{fig:fig 2c}
\end{figure}

The performance of the SC technique is depicted in Fig.~\ref{fig:fig 2c}, where Monte Carlo simulations are compared to that of analytical equations. The probability of success of SC combining technique surpasses the no repetition scenario for both channel conditions. The path-loss only condition reveals a better performance. Additionally, for both no repetition and SC, the signal power declines with an increase in the distance between device and gNB, and so does the probability of success. However for the SC, the performance of devices at large distance shows an improved probability of success due to the increase in the repetitions $N$ configured by the device. Since the number of repetition $N$ is a discrete value, a stepping effect is observed in the probability of success. Note that there is an optimal combination parameters $a$ and $b$ that maximizes the coverage probability, as very low values of $N$ lead to high frame losses and very high $N$ leads to elevated frame collisions. The optimization concept is thoroughly elaborated in~\cite{RepExRepPaper}.

\subsection{Maximal Ratio Combining}\label{MRC}
The aim of this technique is to maximize the resulting SINR by combining the $N$ repetition transmissions. To obtain an accurate analysis, all transmissions in a repetition undergoes separate fading. The received transmissions are added before further signal processing, leading to an overall gain in the SINR. 
Implementing the geometric model as explained in~\ref{geomodel}, for a device at a given location transmitting $N$ repetitions, the received signal at the gNB when using MRC is given as in eq.~(13.17)~\cite{RepExmolish}, 
\begin{align}
    \gamma_\mathrm{MRC}=\sum_{n=1}^{N} \beta_\mathrm{n},
\end{align}
where $\beta_\text{n}$ is the SNR of the each repetition. We extend the case by including interference\cite{RepExMRC,RepExMRC2,RepExMRC3} from the other active devices. We assume the average interference across all $n$ repetitions. Nevertheless, we account for the spatial correlation of $I$ using the repetition profile and~\eqref{agI}.

Thus, the SINR for a device at a distance $r_\text{o}$ and using $N$ repetitions is formulated as follows,
\begin{align}\label{eq21}
      \gamma(N) = \frac{\sum_{n=1}^{N}h_{\mathrm{o}_n}r_o^{-\alpha}} {I+\sigma^2},
\end{align}
where $h_{\mathrm{o}_n}$ is the fading power gain of $n^{th}$ repetition and $I$ is the aggregated interference power as provided in~\eqref{agI}. The probability of success for MRC under various channel losses is evaluated in next subsections.
   
 \subsubsection{MRC with path-loss only}\label{MRC1}
 Under the impact of the path-loss only, from~\eqref{eq21}, the probability of success for $N$ repetitions, conditioned at a given $r_\text{o}$ reduces to, 
 \begin{align}\label{eq10re}
     p_{|r_\mathrm{o}}(N)&=\mathbb{P}({\gamma(N) >  \theta}|r_\mathrm{o},h) \nonumber\\&=   \mathbb{P}\left(\frac{\sum_{n=1}^{N}h_{\mathrm{o}_n}r_\mathrm{o}^{-\alpha} }{I+\sigma^2}>\theta\right)\nonumber\\
    &\overset{(\mathrm{a})}=\mathbb{P}\left(\frac{N(r_\mathrm{o})r_\mathrm{o}^{-\alpha} }{I+\sigma^2}>\theta\right)~,
\end{align}
 where (a) follows from the assumption of fading-less channel i.e., $h_\mathrm{o}$ = 1. 
 Accordingly, the probability of success of $N$ repetitions is given as,
 \begin{align}\label{MrcNoF}
    p_{|r_\mathrm{o}}(N) &=F_I\left(\frac{N(r_\mathrm{o})r_\mathrm{o}^{-\alpha}}{\theta}-\sigma^2\right).
 \end{align}
 $F_I$ is obtained from \eqref{eq6}, where $\mathcal{L}_I(s)$ is used from \eqref{eq8re}. 
 \begin{figure}[t]
    \includegraphics[width=\linewidth]{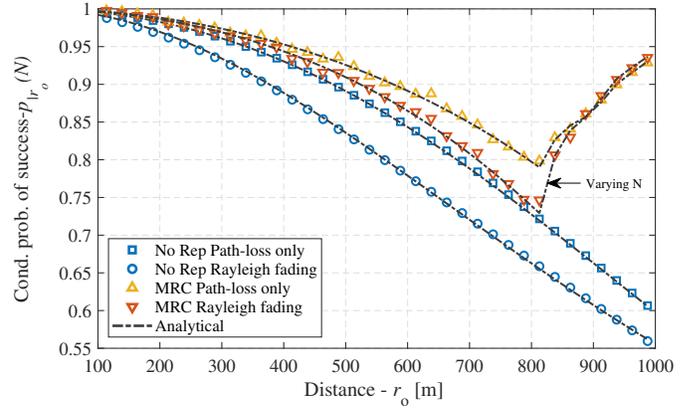}
        \caption{Conditional success probability vs $r_\mathrm{o}$ for no repetition and the two cases of MRC from~\eqref{MrcNoF} and \eqref{eq12}, respectively. The utilized repetition profile parameters are $a = 50$, $b=1050$.}
    \label{fig:fig 4re}
\end{figure}
\subsubsection{MRC with Rayleigh Fading}\label{MRC2}
Here we assume, the serving and the interfering signals are impacted by both
Rayleigh fading and path-loss. Since the fading impacts all the transmissions in $N$ repetitions, it is significant to capture its impact on the resulting SINR due signal combining at the receiver. Utilizing~\eqref{eq21} the success probability for $N$ repetitions, conditioned at $r_\text{o}$ and $h_\text{o}$, is obtained as follows, 
\begin{align}\label{eq11re}
      p_{|r_\mathrm{o},h}(N)&=\mathbb{P}({\gamma(N) >  \theta}|r_\mathrm{o},h) \nonumber\\&=   \mathbb{P}\left(\frac{\sum_{n=1}^{N}h_{\mathrm{o}_n}r_\mathrm{o}^{-\alpha} }{I+\sigma^2}>\theta\right) \nonumber\\ &=  F_{I}\left(\frac{G_\mathrm{o}r_\mathrm{o}^{-\alpha}}{\theta}-\sigma^2\right)~,
\end{align}
where $G_\text{o}$ is the gamma random variable, and is equal to sum of the exponential random variables $h_{\mathrm{o}_1}+h_{\mathrm{o}_2}+..h_{\mathrm{o}_3}$ \cite{RepExMRC}.
Deconditioning on $G_\text{o}$, the success probability for $N$ repetitions is given as, 
\begin{align}\label{eq12}
      p_{|r_\mathrm{o}}(N)=\int_o^\infty F_{I}\left(\frac{u r_\mathrm{o}^{-\alpha}}{\theta}-\sigma^2\right) f_{G_{o}}(u)~\mathrm{d}u~,
\end{align}
where $f_{G_\text{o}}(u)$ is the PDF of the gamma distribution with number of repetitions $N(r_\mathrm{o})$ and scale 1 and is given as follows,
\begin{align}\label{eq12b}
      f_{G_{o}}(u) = \frac{u^{N(r_o)-1}\exp({- u})}{\Gamma\left(N\left(r_o\right)\right)}.
\end{align}
\begin{figure}[t]
    \includegraphics[width=\linewidth]{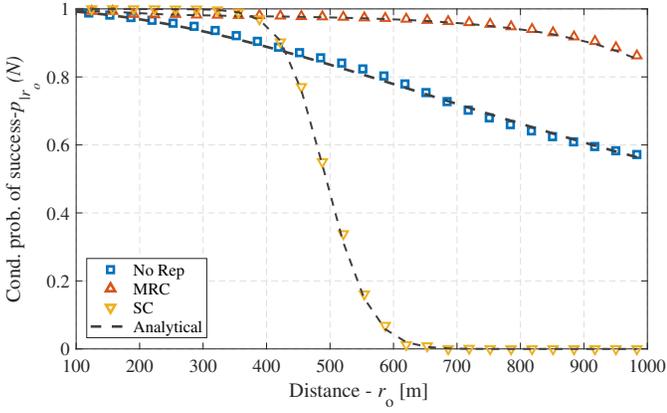}
        \caption{Conditional success probability vs $r_\mathrm{o}$ for no repetition, SC, and MRC, with path-loss and fading, at repetition profile parameters $a = 100$, $b=500$.}
    \label{fig:fig 5re}
\end{figure}
\begin{figure}[t]
    \includegraphics[width=\linewidth]{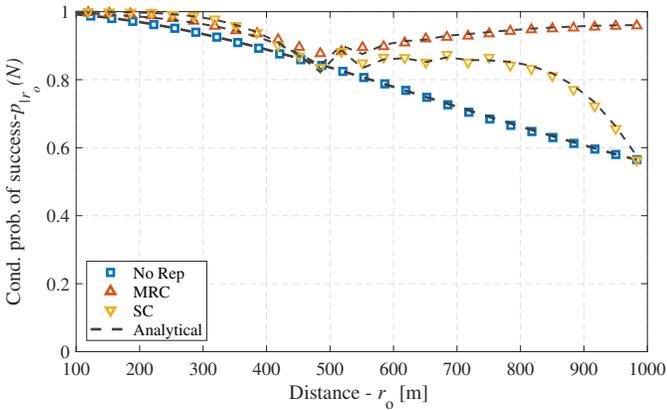}
        \caption{Conditional success probability vs $r_\mathrm{o}$ for no repetition, SC , and MRC, with path-loss and fading, at repetition profile parameters $a = 100$, $b=950$.}
    \label{fig:fig 6re}
\end{figure}
The performance of MRC technique compared to a no repetition scenario is presented in Fig.~\ref{fig:fig 4re}. The output is illustrated for different repetition profile function parameters $a$ and $b$ along with Monte Carlo simulation showing a close match with the theoretical formula. The path-loss only scenario aids in providing the better performance of success probability, and the channel condition with path-loss and Rayleigh fading gives a lower success probability. Furthermore, comparison of both combining techniques with the no repetition scenario is illustrated in Fig.~\ref{fig:fig 5re} and Fig.~\ref{fig:fig 6re}. Fig.~\ref{fig:fig 5re} demonstrates the enhancement provided by MRC over the SC technique at small value of $b$ indicting higher repetition values in a cell. Contrary, at $b=950$, in Fig.~\ref{fig:fig 6re}, the overall repetitions used in a cell reduce and only the devices near the cell edge employ extensive repetition. This reduces the impact of overall interference and hence an adequate performance is obtained. In this case, both techniques show a better performance than a no repetition scenario. As such from a coverage perspective, MRC is beneficial in a cell with using high or low repetitions while SC performs better only at low repetition profile.

On one hand, the execution of the repetition scheme yields an enhanced coverage, however, on the other hand the device is consuming more battery. Since most of the mMTC devices are battery operated and are expected to have a battery life of about 10~years, they rely on efficient energy consumption~\cite{RepExenergy10}. Thus, for the combining schemes showing acceptable coverage performance under various repetition profiles, it is equivalently vital to examine the energy expenditure of an IoT device for both combining schemes at given repetition profiles.
\section{Cell energy wastage due to repetitions} 
Among various operational modes of an IoT device, the ideal and Power Saving Mode (PSM) modes aim at increasing the battery lifetime. During the connected mode, a piece of information is conventionally transmitted once, however, if the device is configured to repeat, the same message is sent over multiple times. The prolonged transmission due to the repetitions significantly impacts the energy consumption of a device. As such, this section focuses on the energy consumption of the device with repeated uplink transmissions. 
\begin{figure}
    \centering
    \includegraphics[width=\linewidth]{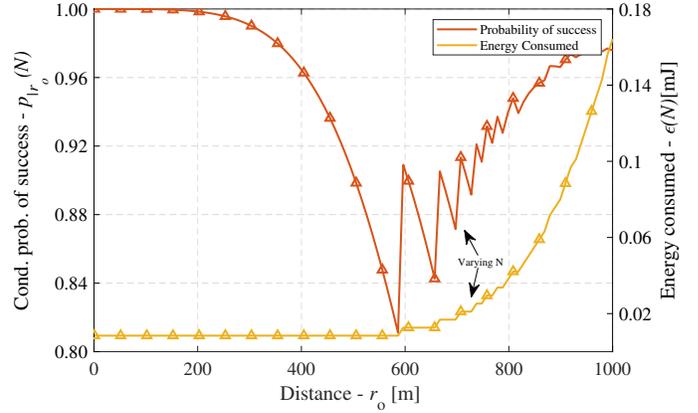}
    \caption{Illustration of the trade-off between probability of success and energy consumption at repetition profile $a=100$ and $b=1050$.}
    \label{fig:EngvsPs}
\end{figure}
We mathematically model the energy consumption of an IoT device during its transmission mode. Assuming an IoT device transmitting a message with time-on-air $T_{m}$ while utilizing the RF transmit power. The energy consumed by the device during this single transmission is given as follows~\cite{OPTZ8935360} ,
\begin{align}\label{eq14re}
\epsilon(1) &=(\eta_\epsilon P_\mathrm{t}+P_\mathrm{O})T_{m},
\end{align}
where $\eta_\epsilon$ is the conversion factor of the power amplifier\cite{OPTZ8935360} and values at least 1, $P_\text{O}$ is the overhead power consumed by the communication module of the device while encoding~\cite{RepExEnergyA}. Further, if the device repeats the transmission $N$ times, it stays active for an extended duration. Thus, the total energy consumed in case of $N$ transmissions is given as follows, 
\begin{equation}\label{eq15}
\epsilon(N) =\epsilon(1) N(r)=
(\eta_\epsilon P_\mathrm{t}+P_\mathrm{O})T_{m} N(r),
\end{equation}
where $N(r)$ is the repetition profile evaluated using~\eqref{eq2re}. As seen in the previous sections, increase in number of transmissions enhance the probability of success, however throughout this process the device consumes more energy. An illustration of this trade-off is shown in Fig.~\ref{fig:EngvsPs}. On one hand, an initial increase in the distance between the device and the gNB reduces the probability of success, however due to the increasing repetitions $N$, an enhancement in probability of success is achieved. On the other hand, using the same repetition profile, more energy is consumed by an IoT device with the increasing values of $N$.
Furthermore, a transmission is deemed successful when at least a single packet out of the $N$ repetitions is delivered successfully. 
Failure in delivery of repetitions leads to error and the energy expended by the device is wasted. 
Thus, the probability of error and the energy consumed aids in evaluating the energy wasted while using the repetition scheme. Therefore, the wasted energy, for an IoT device at distance on $r_\text{o}$ when no repetition is successfully delivered is put forth as, \begin{align}\label{eq17}
\epsilon_{\mathrm{w}}(N)=\epsilon(N) p_e{_{|r_\mathrm{o}}}(N),
\end{align}
\begin{figure}
    \includegraphics[width=\linewidth]{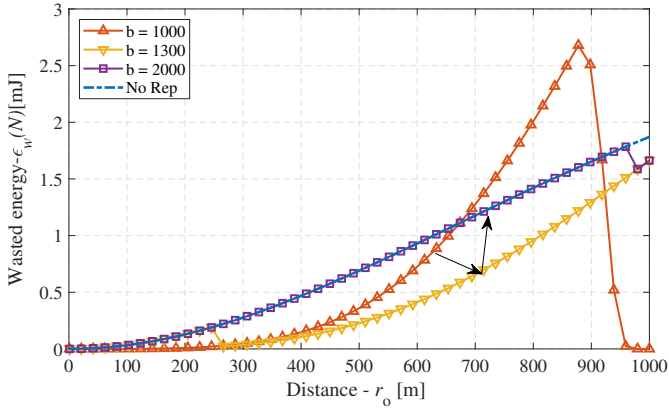}
        \caption{Energy wastage for SC using~\eqref{eq17} vs distance from the gNB, at three different values of repetition profile parameter $b$ and a constant $a$ = 25.}
    \label{fig:fig9}
\end{figure}where $p_e{_{|r_\mathrm{o}}(N)}$ is given from \eqref{pe}.
The wasted energy is a function of the i) number of repetitions $N$, and ii) the probability of error, which depends on the distance of the device from gNB and as well as on the repetition profile parameters $a$ and $b$.
Thus, to examine the impact of various repetition profile parameters on energy dissipation we assess the average wasted energy across the cell over the distribution of the distances of devices to the gNB. The expected wasted energy in a cell at various repetition profile parameters is obtained by deconditioned~\eqref{eq17} on $r_\text{o}$ given as,
\begin{align}\label{eq18}
\overline{\epsilon}_\mathrm{c}=\mathbb{E}\left[\epsilon_{{\mathrm{waste}}(N)}\right]=\int_{0}^{R}\epsilon_{{\mathrm{w}}(N)} f_r(r_\mathrm{o})  ~\mathrm{d}r~,
\end{align}
where $f_r(r_\mathrm{o})$ is the distance probability density function (pdf) and is given as, 
\begin{align}\label{eq19}
     f_r(r_\mathrm{o})=\frac{\mathrm{d}}{\mathrm{d}r}\left[\frac{\text{avg. \# of points in $r_\mathrm{o}$} }{\text{avg. \# of points in $R_\mathrm{c}$}}\right] =\frac{2\pi r_\mathrm{o}D(r) }{{\eta(R_\mathrm{c})}}~,
\end{align}
and $\eta$ is the cell constant given by~eq~(6) in~\cite{RepExRepPaper}. The wasted energy for SC as a function of distance at three repetition profiles is shown in Fig.~\ref{fig:fig9}. For all repetition profiles including the no repetition scenario, the gradual increase in the distance leads to an increase in the probability of error and wasted energy. Nevertheless, the distinction between the repetition profiles is due to the parameter $b$ which impacts the overall values of $N$ in a cell. For the first profile at $b=1000$, the values of $N$ enhance near the cell edge, as such the likelihood of error reduces, leading to drop in energy wasted. The infliction point in the figure depicts a point where the decline in probability of error due to enhancing $N$ overshadows the energy consumption. For the second profile at $b=1300$ the overall values of $N$ in a cell are less compared to the first profile. At these values a high success probability is achieved, and hence minimum energy is wasted.  It is interesting to note that in this case the wasted energy is less than that of the no repetition scenario. For the third repetition profile at $b=2000$ the value of $N$ reduces to 1, thereby no enhancement in probability of success is achieved. This phenomenon occurs due to the reason that on one hand, the large values of $N$ lead to network collisions that results in less success probability in SC, and on the other hand, the minimum value of $N$ entails no repetition. Thus, there exists an optimal repetition profile in between, with values of $N$ that lead to a high success probability and minimum wasted energy. 
The wasted energy for MRC at similar repetition profiles is shown in Fig.~\ref{fig:fig10}. However, in MRC, the wasted energy does not reduce below that of no repetition scenario for any repetition profiles.
 As such, the cell average wasted energy is evaluated only for SC at various $b$ as shown in Fig.~\ref{fig:fig11}. The influence of the different values of the repetition profile parameter $b$ is clearly visible on the average cell wasted energy. It is seen that, the average wasted energy across the cell is less than that of no repetition scenario for various values of parameter $b$.
\begin{figure}[t]
    \includegraphics[width=\linewidth]{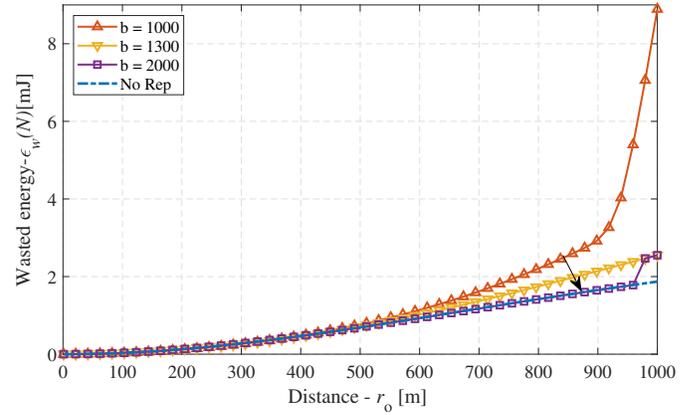}
      \caption{Energy wastage obtained for MRC using~\eqref{eq17} versus distance from the gNB at three different values of repetition profile parameter $b$ and a constant $a$ = 25.}
    \label{fig:fig10}
\end{figure}

\begin{figure}
    \includegraphics[width=\linewidth]{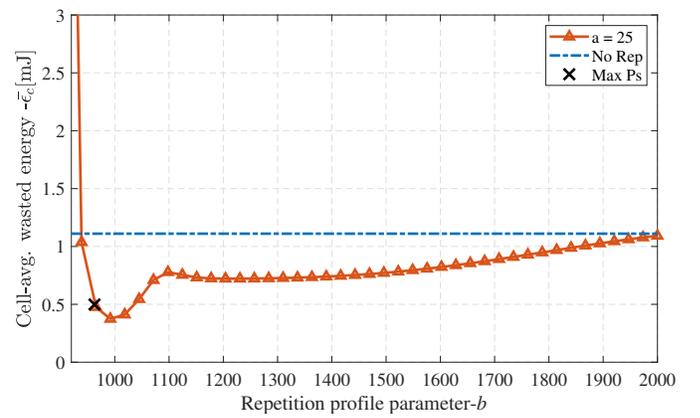}
        \caption{Average energy wasted across a cell using \eqref{eq18} for SC vs repetition profile parameter $b$ in comparison to no repetition scenario.}
    \label{fig:fig11}
\end{figure}
\begin{figure}[ht]
    \includegraphics[width=\linewidth]{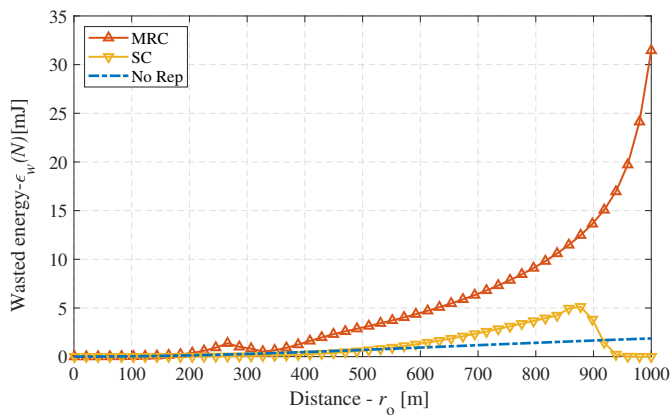}
      \caption{Comparison of energy wastage between MRC and SC for high success rate (97\%) obtained at repetition profile parameters $a = 25$, $b = 500$ for MRC, and $a = 25$, $b = 960$ for SC.}
    \label{fig:fig12}
\end{figure}

Furthermore, a comparison of the wasted energy at a high coverage probability (97\%) for both combining schemes is illustrated in Fig.~\ref{fig:fig12}. On one hand, this coverage probability for MRC is achieved at large values of $N$ at a repetition profile with parameters $b=500$ and $a=25$. However, the energy wastage is higher compared to that of SC and no repetition, as demonstrated. 
On the other hand, less energy is wasted while providing the same service by SC at lower values $N$ for repetition profile parameters $b=960$ and $a=25$. Hence, for a given coverage, SC is more energy-efficient than that of MRC. As a result, application of combining techniques is subject to requirements of coverage and energy efficiency.  
\section{Conclusion}
In this paper an analysis of two signal combining techniques, SC and MRC is performed for IoT repetition scheme in terms of probability of success and energy efficiency. The employed repetition mechanism is based on the distance of the device from the gNB and the duty cycle. Both techniques are analyzed for (i) path-loss only scenario and (ii) path-loss with fading scenario. All cases are compared to that of the no repetition scenario. The evaluation reveals that both techniques respond differently under various optimal repetition profiles. In terms of coverage, MRC outperforms SC at high and low repetition rates in a cell. Comparatively, SC shows an improved coverage only for a low repetition profile. Furthermore, a mathematical model to capture the prevalent energy wastage due to repetition failure was also presented. The model demonstrates that SC is more energy-efficient as compared to MRC and reveals the point of lowest energy wastage when a given repetition profile is realized in a cell.

Therefore, in an IoT cell utilizing extensive repetition values, an enhanced coverage can be achieved by using the MRC technique at the receiver. Moreover, if low repetition values are implemented, both combining techniques are suitable. Nevertheless, from an energy outlook, SC is a more appropriate combining technique for energy-constrained environment.
\bibliographystyle{IEEEtran}
\bibliography{rep_tgcn} \vspace{-1.4cm}
\begin{IEEEbiography}
[{\includegraphics[width=1in,height=1.25in,clip,keepaspectratio]{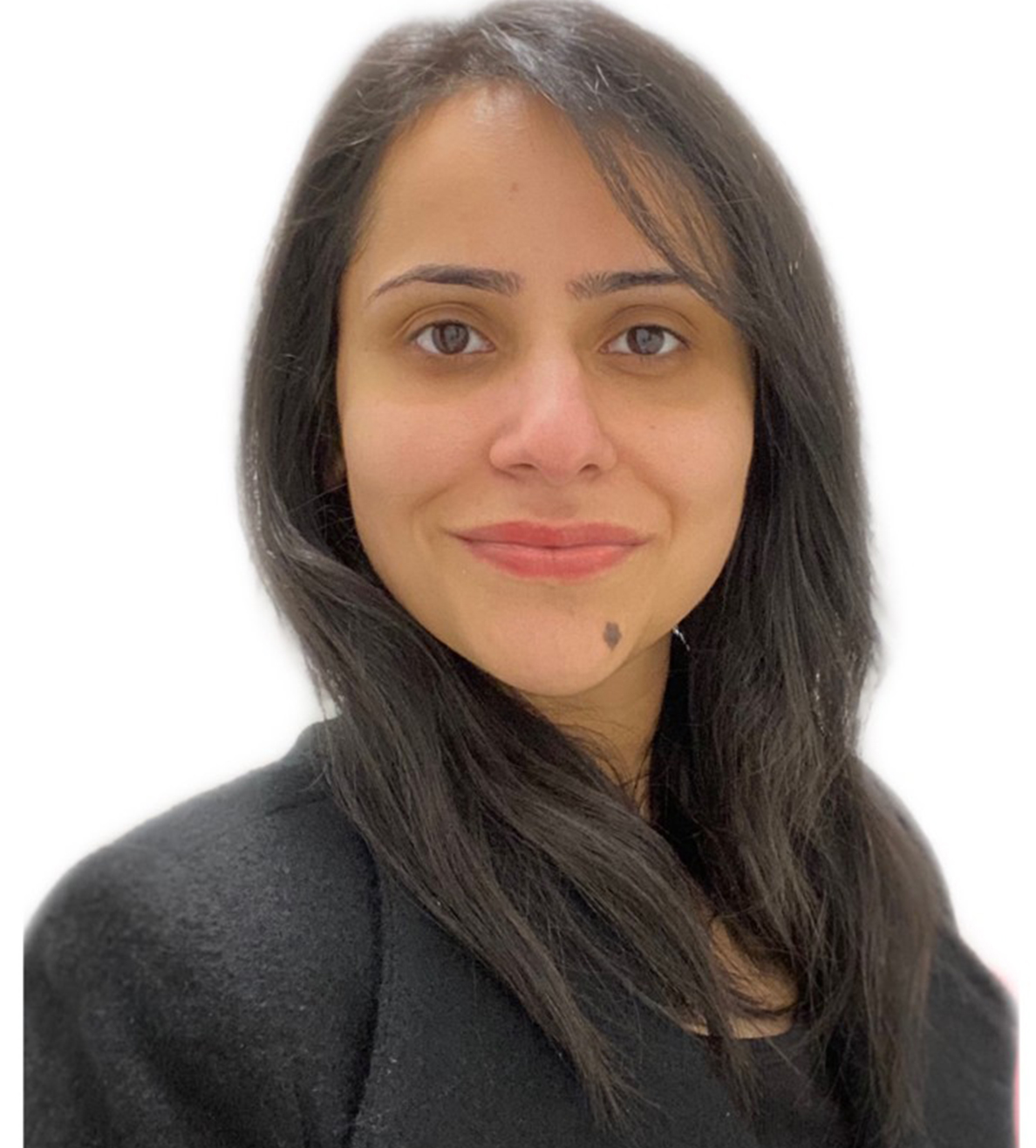}}]{Dr Bisma Manzoor} (PhD, MEng, BEng) received the PhD degree in Electronic and Electrical Engineering, with specialization in wirelesses communication systems and technologies, from the School of Engineering, RMIT University, Melbourne, Australia, in 2021. She received her Master’s degree in Telecommunication and Network Engineering from RMIT University, Melbourne Australia, in 2017. Dr Manzoor currently works as signal processing engineer at Fleet Space Technologies, Australia. Her research interests include modeling and optimization for the cellular Internet of Things, energy efficiency in wireless networks, Internet of Things over satellite, and Stochastic Geometry. 
\end{IEEEbiography} 
\vspace{-1cm}
\begin{IEEEbiography}[{\includegraphics[width=1in,height=1.25in,clip,keepaspectratio]{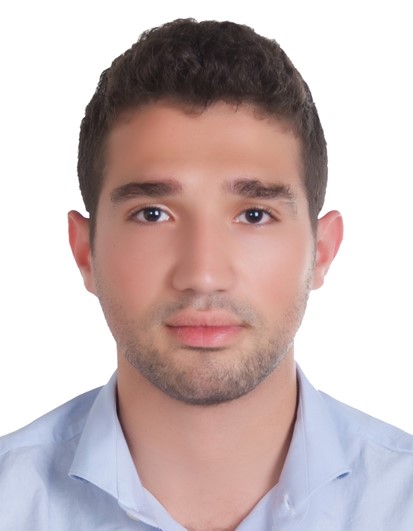}}]{Dr Bassel Al Homssi} (PhD, EIT, MSEng, BSEng) is a Research Fellow at both RMIT University and Deakin University, Melbourne, Australia. Dr Al Homssi received his Bachelor of Science and Master of Science in Electrical and Electronic Engineering from the American University of Sharjah, Sharjah, United Arab Emirates in 2014 and 2016 respectively. He also received his Engineer in Training (EIT) certification from NCEES in 2014 and also received his Master in Telecommunication from the University of Melbourne, Melbourne, Australia in 2017. Dr Al Homssi received his PhD degree in 2020 from RMIT University and was the lead network designer for North Melbourne Smart Cities Network. He is currently working with the local Australian government on Smart Cities in Melbourne, Australia. Dr Al Homssi is currently working on multiple research projects related to satellite networks and MIMO communications for next generation wireless systems.
\end{IEEEbiography} 
\begin{IEEEbiography}
[{\includegraphics[width=1in,height=1.25in,clip,keepaspectratio]{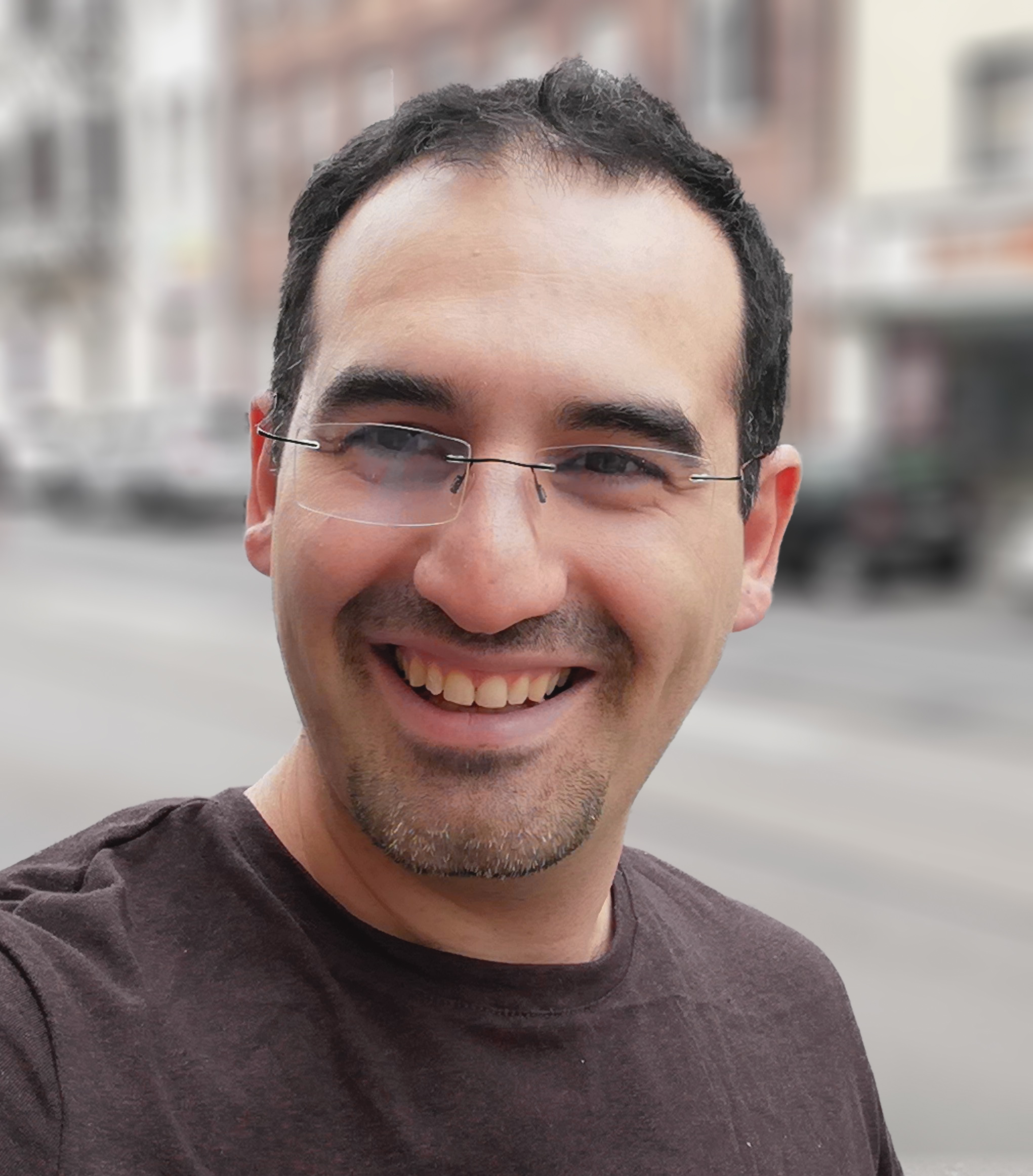}}]{Dr Akram Al-Hourani} (PhD, BEng, MBA, SMIEE), is a Senior lecturer and Telecommunication Program Manager at the School of Engineering, RMIT University. Before joining RMIT in 2016, he has been with The University of Melbourne as a Research Fellow. Dr Al-Hourani obtained his PhD in 2016 from RMIT University. Prior joining academia, Dr Al-Hourani had extensively worked in the ICT industry as an R\&D engineer, radio network planning engineer and then as an ICT program manager for several projects spanning over different technologies; including mobile networks deployment, satellite networks, and railway ICT systems. The total value of industry projects directed by Dr Al-Hourani exceeded \$190m. He has extensive industry / government engagement as a chief investigator in multiple research projects related to The Internet-of-Things (IoT), Smart Cities, Satellite / Wireless Communications. As a Lead Chief Investigator, he provided key expertise in delivering the largest open (Internet-of-Things) network in Victoria in collaboration with 5 local governments “Northern Melbourne Smart Cities Network”, this project attracted two prestigious industry awards from the Municipal Association of Victoria (MAV) "Technology Awards for Excellence 2020", and the IoT Alliance Australia (IoTAA) "Smart Cities Award for 2020". He published more than 75 journal articles and conference proceedings, including 3 book chapters, Dr Al-Hourani has won the IEEE Sensors Council Paper Award for his contribution in hand-gesture recognition using neural networks. He is currently listed in Stanford University's list of top 2\% scientists in both Career-Long list and Single Year Impact list for 2020. Dr Al-Hourani is currently an Associate Editor in IEEE Transactions of Aerospace and Electronics, and in Frontiers in Space Technologies and an Editor in MDPI Remote Sensing Journal. His current research interests include the Satellite Communications, Radars Systems, UAV Communication Systems, Signal Processing / Machine Learning, and Stochastic Geometry.
\end{IEEEbiography}\vfill
\vfill \vfill \vfill \vfill \vfill \vfill \vfill \vfill \vfill \vfill \vfill \vfill \vfill \vfill \vfill \vfill \vfill \vfill 
\vfill 
\vfill \vfill \vfill \vfill \vfill \vfill \vfill \vfill \vfill \vfill \vfill \vfill \vfill \vfill \vfill \vfill \vfill \vfill 
\vfill 
\vfill \vfill \vfill \vfill \vfill \vfill \vfill \vfill \vfill \vfill \vfill \vfill \vfill \vfill \vfill \vfill \vfill \vfill \vfill \vfill \vfill \vfill \vfill \vfill \vfill \vfill \vfill \vfill \vfill \vfill \vfill \vfill \vfill \vfill \vfill \vfill 
\vfill 
\vfill \vfill \vfill \vfill \vfill \vfill \vfill \vfill \vfill \vfill \vfill \vfill \vfill \vfill \vfill \vfill \vfill \vfill 
\vfill 
\vfill \vfill \vfill \vfill \vfill \vfill \vfill \vfill \vfill \vfill \vfill \vfill \vfill \vfill \vfill \vfill \vfill \vfill \vfill \vfill \vfill \vfill \vfill \vfill \vfill \vfill \vfill \vfill \vfill \vfill \vfill \vfill \vfill \vfill \vfill \vfill 
\vfill 
\vfill \vfill \vfill \vfill \vfill \vfill \vfill \vfill \vfill \vfill \vfill \vfill \vfill \vfill \vfill \vfill \vfill \vfill 
\vfill 
\vfill \vfill \vfill \vfill \vfill \vfill \vfill \vfill \vfill \vfill \vfill \vfill \vfill \vfill \vfill \vfill \vfill \vfill \vfill \vfill \vfill \vfill \vfill \vfill \vfill \vfill \vfill \vfill \vfill \vfill \vfill \vfill \vfill \vfill \vfill \vfill 
\vfill 
\vfill \vfill \vfill \vfill \vfill \vfill \vfill \vfill \vfill \vfill \vfill \vfill \vfill \vfill \vfill \vfill \vfill \vfill 
\vfill 
\vfill \vfill \vfill \vfill \vfill \vfill \vfill \vfill \vfill \vfill \vfill \vfill \vfill \vfill \vfill \vfill \vfill \vfill \vfill \vfill \vfill \vfill \vfill \vfill \vfill \vfill \vfill \vfill \vfill \vfill \vfill \vfill \vfill \vfill \vfill \vfill 
\vfill 
\vfill \vfill \vfill \vfill \vfill \vfill \vfill \vfill \vfill \vfill \vfill \vfill \vfill \vfill \vfill \vfill \vfill \vfill 
\vfill 
\vfill \vfill \vfill \vfill \vfill \vfill \vfill \vfill \vfill \vfill \vfill \vfill \vfill \vfill \vfill \vfill \vfill \vfill \vfill \vfill \vfill \vfill \vfill \vfill \vfill \vfill \vfill \vfill \vfill \vfill \vfill \vfill \vfill \vfill \vfill \vfill 
\vfill 
\vfill \vfill \vfill \vfill \vfill \vfill \vfill \vfill \vfill \vfill \vfill \vfill \vfill \vfill \vfill \vfill \vfill \vfill 
\vfill 
\vfill \vfill \vfill \vfill \vfill \vfill \vfill \vfill \vfill \vfill \vfill \vfill \vfill \vfill \vfill \vfill \vfill \vfill \vfill \vfill \vfill \vfill \vfill \vfill \vfill \vfill \vfill \vfill \vfill \vfill \vfill \vfill \vfill \vfill \vfill \vfill \vfill \vfill \vfill \vfill \vfill \vfill \vfill \vfill \vfill \vfill \vfill 
\vfill 
\vfill \vfill \vfill \vfill \vfill \vfill \vfill \vfill \vfill \vfill \vfill \vfill \vfill \vfill \vfill \vfill \vfill \vfill 
\vfill 
\vfill \vfill \vfill \vfill \vfill \vfill \vfill \vfill \vfill \vfill \vfill \vfill\vfill \vfill \vfill \vfill \vfill \vfill \vfill \vfill \vfill \vfill \vfill \vfill \vfill \vfill \vfill \vfill \vfill \vfill \vfill \vfill \vfill \vfill \vfill \vfill \vfill \vfill \vfill \vfill \vfill 
\vfill 
\vfill \vfill \vfill \vfill \vfill \vfill \vfill \vfill \vfill \vfill \vfill \vfill \vfill \vfill \vfill \vfill \vfill \vfill 
\vfill 
\vfill \vfill \vfill \vfill \vfill \vfill \vfill \vfill \vfill \vfill \vfill \vfill\vfill \vfill \vfill \vfill \vfill \vfill \vfill \vfill \vfill \vfill \vfill \vfill \vfill \vfill \vfill \vfill \vfill \vfill \vfill \vfill \vfill \vfill \vfill \vfill \vfill \vfill \vfill \vfill \vfill 
\vfill 
\vfill \vfill \vfill \vfill \vfill \vfill \vfill \vfill \vfill \vfill \vfill \vfill \vfill \vfill \vfill \vfill \vfill \vfill 
\vfill 
\vfill \vfill \vfill \vfill \vfill \vfill \vfill \vfill \vfill \vfill \vfill \vfill\vfill \vfill \vfill \vfill \vfill \vfill \vfill \vfill \vfill \vfill \vfill \vfill \vfill \vfill \vfill \vfill \vfill \vfill \vfill \vfill \vfill \vfill \vfill \vfill \vfill \vfill \vfill \vfill \vfill 
\vfill 
\vfill \vfill \vfill \vfill \vfill \vfill \vfill \vfill \vfill \vfill \vfill \vfill \vfill \vfill \vfill \vfill \vfill \vfill 
\vfill 
\vfill \vfill \vfill \vfill \vfill \vfill \vfill \vfill \vfill \vfill \vfill \vfill\vfill \vfill \vfill \vfill \vfill \vfill \vfill \vfill \vfill \vfill \vfill \vfill \vfill \vfill \vfill \vfill \vfill \vfill \vfill \vfill \vfill \vfill \vfill \vfill \vfill \vfill \vfill \vfill \vfill 
\vfill 
\vfill \vfill \vfill \vfill \vfill \vfill \vfill \vfill \vfill \vfill \vfill \vfill \vfill \vfill \vfill \vfill \vfill \vfill 
\vfill 
\vfill \vfill \vfill \vfill \vfill \vfill \vfill \vfill \vfill \vfill \vfill \vfill\vfill \vfill \vfill \vfill \vfill \vfill \vfill \vfill \vfill \vfill \vfill \vfill \vfill \vfill \vfill \vfill \vfill \vfill \vfill \vfill \vfill \vfill \vfill \vfill \vfill \vfill \vfill \vfill \vfill 
\vfill 
\vfill \vfill \vfill \vfill \vfill \vfill \vfill \vfill \vfill \vfill \vfill \vfill \vfill \vfill \vfill \vfill \vfill \vfill 
\vfill 
\vfill \vfill \vfill \vfill \vfill \vfill \vfill \vfill \vfill \vfill \vfill \vfill
\end{document}